\documentclass[11pt, a4paper, twoside]{article}
\usepackage{geometry}
\usepackage{amsmath,amsfonts,amssymb}
\usepackage[utf8]{inputenc}
\usepackage[english]{babel}
\usepackage[runin]{abstract}
\usepackage{titling}
\usepackage{booktabs}
\usepackage{fancyhdr}
\usepackage{helvet}
\usepackage[dvipsnames]{xcolor}
\usepackage{csquotes}
\usepackage{graphicx}
\usepackage{hyperref}
\usepackage{breakurl}
\usepackage{threeparttable}
\usepackage{multirow}
\usepackage{caption}
\usepackage{subcaption}
\usepackage{parskip}
\usepackage{etoolbox}
\usepackage[
    backend=biber,
    style=alphabetic,
    giveninits=true,
    sorting=nyt,
    hyperref=auto,
]{biblatex}

\hypersetup{
    colorlinks,
    urlcolor={blue!80!black},
    citecolor={blue!50!black},
    linkcolor={red!50!black},
    breaklinks=true,  
}

\date{}

\makeatletter

\def\fulltitle#1{\def\@fulltitle{#1}}
\def\runningtitle#1{\def\@runningtitle{#1}}
\def\runningauthor#1{\def\@runningauthor{#1}}
\def\affiliation#1{\def\@affiliation{#1}}
\def\department#1{\def\@department{#1}}
\def\memoid#1{\def\@memoid{#1}}
\def\theyear#1{\def\@theyear{#1}}
\def\mydate#1{\def\@mydate{#1}}

\runningtitle{How to write an effective report} 
\author{Your name \and John Doe \and Jane Doe} 
\runningauthor{Your name et al.} 
\affiliation{Your University} 
\department{Your Department} 
\memoid{AI Memo 777} 
\theyear{2018} 
\mydate{August 08, 2018} 

\def\displaymydate{\@mydate}
\def\displaytheyear{\@theyear}
\def\displaymemoid{\@memoid}
\def\displaydepartment{\@department}
\def\displayaffiliation{\@affiliation}
\def\displayrunningauthor{\@runningauthor}
\def\displayrunningtitle{\@runningtitle}

\makeatother

\makeatletter
\patchcmd{\@zfancyhead}{\fancy@reset}{\f@nch@reset}{}{}
\patchcmd{\@set@em@up}{\f@ncyolh}{\f@nch@olh}{}{}
\patchcmd{\@set@em@up}{\f@ncyolh}{\f@nch@olh}{}{}
\patchcmd{\@set@em@up}{\f@ncyorh}{\f@nch@orh}{}{}
\makeatother

\fancypagestyle{firstpage}
{
    \fancyhf{}
    \fancyhead{%
        \begin{center}
            \textsc{\displayaffiliation}\\
            \textsc{\displaydepartment}
        \end{center}
        \vspace*{0.5in}
        \displaymemoid
        \hfill
        \displaymydate
    }
    \fancyfoot[L]{\footnotesize \textbf{\textsc{\displayaffiliation} -- \displaytheyear}}
}

\providecommand{\keywords}[1]{\noindent \textbf{Keywords:} #1} 

\abslabeldelim{:} 
\pretitle{\Large \begin{center}}
\posttitle{\\[1ex] \Large \textbf{by}\\[5ex]\end{center}}
\title{\textbf{Visualizing Multimodality in Combinatorial Search Landscapes}} 
\runningtitle{Visualizing Multimodal Search Landscapes} 
\author{Xavier F. C. Sánchez-Díaz \and Ole Jakob Mengshoel} 
\runningauthor{Sánchez-Díaz and Mengshoel} 
\affiliation{Norwegian University of Science and Technology} 
\department{Department of Computer Science} 
\memoid{NAIS 2025} 
\theyear{2025} 
\mydate{June 18, 2025} 

\begin{document}
\maketitle

\begin{abstract}
  This work walks through different visualization techniques for combinatorial search landscapes, focusing on multimodality.
  We discuss different techniques from the landscape analysis literature, and how they can be combined to provide a more comprehensive view of the search landscape.
  We also include examples and discuss relevant work to show how others have used these techniques in practice, based on the geometric and aesthetic elements of the Grammar of Graphics.
  We conclude that there is no free lunch in visualization, and provide recommendations for future work as there are several paths to continue the work in this field.

\end{abstract}

\vspace{2.5cm}

{\footnotesize
    \noindent
    Poster presented at the 2025 Symposium of the Norwegian Artificial Intelligence Society (NAIS 2025).\\
    June 18th \& 19th, Tromsø, Norway.
}

\thispagestyle{firstpage}

\pagebreak


\newgeometry{top=1.5in,
inner=1in,
outer=0.7in,
bottom=1.5in,
asymmetric,
bindingoffset=0in,
marginparwidth=0in} 
\fancyfootoffset{0pt}

\section{Introduction}

\textbf{Context.} Heuristic optimization algorithms are widely used to solve complex problems in engineering, industry, and science.
Regardless of their representation, we aim to find solutions that are optimal inside the set of all possible solutions (or what we call the \textit{search space}).
The \textit{search landscape} metaphor is a way to ``visualize'' the search space of an optimization problem~\cite{wrightRolesMutationInbreeding1932}, and how we exactly ``see'' this landscape is determined by the plotting techniques we use.
Combinatorial problems in AI are abundant. From pseudo-Boolean problems like model finding~\cite{yolcu19Learning}, sentence summarization~\cite{schumann2020discrete} and feature selection 
\cite{mengshoel2021stochastic, sanchez-diazRegularizedFeatureSelection2024}, to other classical problems like the Traveling Salesman~\cite{boeseNewAdaptiveMultistart1994}, routing~\cite{zhangHybridAlgorithmVehicle2017}, packing problems~\cite{sanchez-diazFeatureIndependentHyperHeuristicApproach2021, sanchez-diazPreliminaryStudyFeatureindependent2020}, scheduling~\cite{lara-cardenasImprovingHyperheuristicPerformance2019} and assignment~\cite{thomsonEntropySearchTrajectories2024}.
Furthermore, quite often we are interested in multiple solutions---multimodality in these problems is common, but its existence is not always acknowledged.
Recent conferences on evolutionary computation had just a few papers on multimodality (or problems with multiple optima), while the majority of the works focused on multi-objective optimization, benchmarking, and genetic programming.\footnote{See \url{https://dl.acm.org/doi/proceedings/10.1145/3638529} and \url{https://link.springer.com/book/10.1007/978-3-031-70085-9}}

\noindent \textbf{Challenges.} How do we get the most information from static visualizations of a combinatorial search landscape?
Combinatorial problems are discrete, and the notions of order and continuity might not be well defined~\cite{reidysCombinatorialLandscapes2002}.
Thus, defining a neighborhood is crucial for calculating locality~\cite{sanchez-diaz_estimating_2024}. 
An interesting idea is to create composite visualizations. However, it is not trivial when an optimization problem contains multiple solutions of interest and the problem has high dimensionality~\cite{massonVisualizingPseudoBooleanFunctions2025, mersmannExploratoryLandscapeAnalysis2011}. Moreover, oftentimes the number of optimal solutions is large, and we might be interested in finding as many as possible~\cite{liSeekingMultipleSolutions2017, sanchez-diaz_estimating_2024}.

\noindent \textbf{Contributions.} In this work, we get an overview of several visualization techniques for combinatorial search landscapes from an information design perspective. We focus on multimodality, and identify their strengths and limitations.
Additionally, we propose a simple framework for combining these visualizations on the basis of their aesthetic attributes using the Grammar of Graphics~\cite{wickhamPracticalToolsExploring2008, wilkinsonGrammarGraphics2005}.


\section{Background}
\label{sec:notation}

\subsection{Fitness Landscapes}

This paper is aimed at researchers with a moderate understanding of optimization but who are not necessarily experts in search landscape analysis.
We now introduce relevant concepts and notation.

We define a search (or fitness) landscape as a tuple $\mathfrak{L} = (\mathcal{X}, f, \mathcal{N})$, where $\mathcal{X}$ is the search space, $f$ is \textit{a generic} fitness\footnote{We use the term fitness as it is compared to the ability to survive of an individual in the context of biological evolution.} function and $\mathcal{N}$ is the neighborhood or any notion of accessibility~\cite{ochoaLandscapeAnalysisOptimisation2024}.
By \textit{generic function} we mean that $f$ could be either a real-valued function, or a pseudo-Boolean function of any arity that we aim to optimize.
For the sake of simplicity, we assume, without loss of generality, that we want to minimize $f$.
The fitness function is \textit{navigated} using a given optimization algorithm, and thus $\mathfrak{L}$ is algorithm-dependent, which in turn determines how the neighborhood $\mathcal{N}$ is constructed.

The search space $\mathcal{X}$ is a set of solutions, which in our case is either a set of binary strings, permutations, or any other combinatorial object~\cite{reidysCombinatorialLandscapes2002}.
We use $\boldsymbol{b}$ to denote a solution in $\mathcal{X}$, as a solution is not necessarily a binary string.

We define a global optimum, denoted as $\boldsymbol{b}^*$, as the optimal solution, i.e., $f(\boldsymbol{b}^*) \leq f(\boldsymbol{b})$ for all $\boldsymbol{b} \in \mathcal{X}$.
A local optimum is a solution $\boldsymbol{b}^+ \in \mathcal{X}$ such that $f(\boldsymbol{b}^+) \leq f(\boldsymbol{b})$ for all $\boldsymbol{b} \in \mathcal{N}(\boldsymbol{b}^+)$, where $\mathcal{N}(\boldsymbol{b}^+)$ is the neighborhood of $\boldsymbol{b}^+$. The set of all local optima in a given landscape $\mathfrak{L}$ is represented with $\mathcal{L}$, and its cardinality (or size) as $|\mathcal{L}|$.

The neighborhood $\mathcal{N}$ of a solution $\boldsymbol{b}$ is the set of solutions that are \textit{close} to $\boldsymbol{b}$.
We define this \textit{closeness} using a distance metric $d$.
For example, in the case of binary strings, the Hamming distance $d_H$ is a common choice for $d$, and the neighborhood of $\boldsymbol{b}$ is the set of all solutions that differ from $\boldsymbol{b}$ in one bit: $\mathcal{N}(\boldsymbol{b}) = \left\{\boldsymbol{b}' \in \mathcal{X} \mid d_H(\boldsymbol{b}, \boldsymbol{b}') = 1\right\}$.
This neighborhood is usually employed in local search, genetic algorithms, and many other solvers that operate with bitstrings.
In this work we do not focus on any specific neighborhood nor distance metric, but use these concepts to understand a search landscape.

Finally, throughout this work, we use several terms to refer to landscape features that resemble the topography of a physical landscape: \textit{peaks}, \textit{valleys}, \textit{plateaus}, \textit{funnels}, and \textit{basins}.
We do not define these terms formally, as their physical analogies should suffice to understand their meaning.

\subsection{Visualization and the Grammar of Graphics}

The Grammar of Graphics is a framework for building graphics, based on the idea of \textit{layering} the different semantic elements of a plot, and then mapping these elements to data.
Originally proposed by Wilkinson~\cite{wilkinsonGrammarGraphics2005}, the Grammar of Graphics has been implemented in software libraries and several programming languages, like R~\cite{wickhamPracticalToolsExploring2008} and Julia~\cite{MakieOrgAlgebraOfGraphicsjl2024}.

From the Grammar of Graphics, we highlight two important concepts: \textit{aesthetics} and \textit{geometries}.
Geometries are the ``physical'' elements of a plot: points, lines, bars, etc. These geometrical objects (which we refer to as \textit{geoms}) are then mapped to the data using aesthetics elements.
Aesthetics, on the other hand,  are the visual properties of the geometries, like color, size, shape and position. See Table~\ref{tab:aesthetics} for a list of common aesthetics on different mediums.

\begin{table}[tb]
    \centering
    \caption{Aesthetic attributes in the Grammar of Graphics. Adapted from Table 10.1 in~\cite{wilkinsonGrammarGraphics2005}}
    \label{tab:aesthetics}
    \begin{tabular}{@{}lllll@{}}
    \toprule
    \multicolumn{1}{c}{\textbf{Form}} & \multicolumn{1}{c}{\textbf{Surface}} & \multicolumn{1}{c}{\textbf{Motion}} & \multicolumn{1}{c}{\textbf{Sound}} & \multicolumn{1}{c}{\textbf{Text}} \\ \midrule
    \begin{tabular}[c]{@{}l@{}}position\\ size\\ shape\\ rotation\\ resolution\end{tabular} & \begin{tabular}[c]{@{}l@{}}color:\\ \quad hue\\ \quad brightness\\ \quad saturation\\ texture:\\ \quad pattern\\ \quad granularity\\ \quad orientation\\ blur\\ transparency\end{tabular} & \begin{tabular}[c]{@{}l@{}}direction\\ speed\\ acceleration\end{tabular} & \begin{tabular}[c]{@{}l@{}}tone\\ volume\\ rhythm\\ voice\end{tabular} & label \\ \bottomrule
    \end{tabular}
    
\end{table}

The notion of aesthetic elements and geometric objects is useful when designing visualizations for search landscapes, as we can use these elements to highlight different aspects of the landscape~\cite{gleicherConsiderationsVisualizingComparison2018, javedExploringDesignSpace2012}.
Most importantly, the main advantage is the natural \textit{association} between aesthetics elements and the characteristics of the landscape that can be visualized using them.
For example, some visualizations employ color to represent the fitness of a solution, or use the size of a node to communicate the size of a basin of attraction.
Using this framework, one can design and analyze visualizations that are both informative and aesthetically pleasing.

\subsection{Landscapes and Visualization}
\label{sec:landscapes}

From Wright's conception of the fitness landscape~\cite{wrightRolesMutationInbreeding1932} we have inherited the notion of \textit{peaks} and \textit{valleys} in the search space.
Since then, the \textit{visual features} of a search landscape have been discussed in detail~\cite{pitzerComprehensiveSurveyFitness2012, reevesFitnessLandscapes2014}.
We now have the notion of \textit{ridges}, \textit{plateaus} and \textit{basins}, and have identified more global structures like \textit{funnels} and \textit{canals}~\cite{malanRecentAdvancesLandscape2021, ochoaRecentAdvancesFitness2019, ochoaLandscapeAnalysisOptimisation2024}.

\begin{figure}
    \centering
    \begin{subfigure}[b]{0.33\textwidth}
        \includegraphics[width=\textwidth]{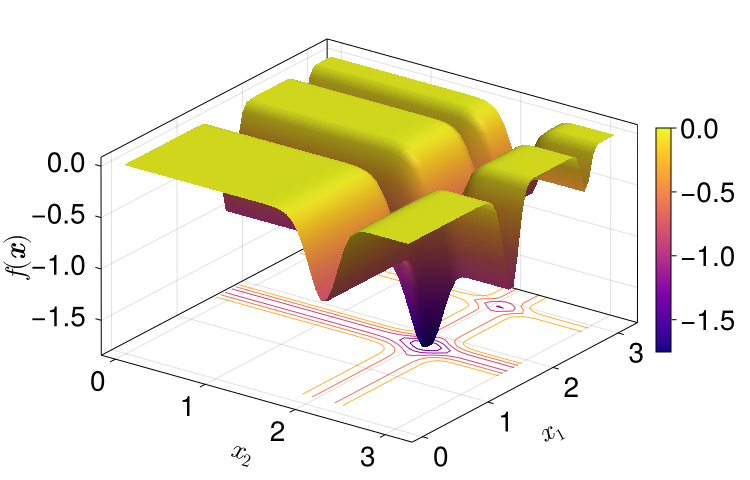}
        \caption{Michalewicz function}
    \end{subfigure}%
    \begin{subfigure}[b]{0.33\textwidth}
        \includegraphics[width=\textwidth]{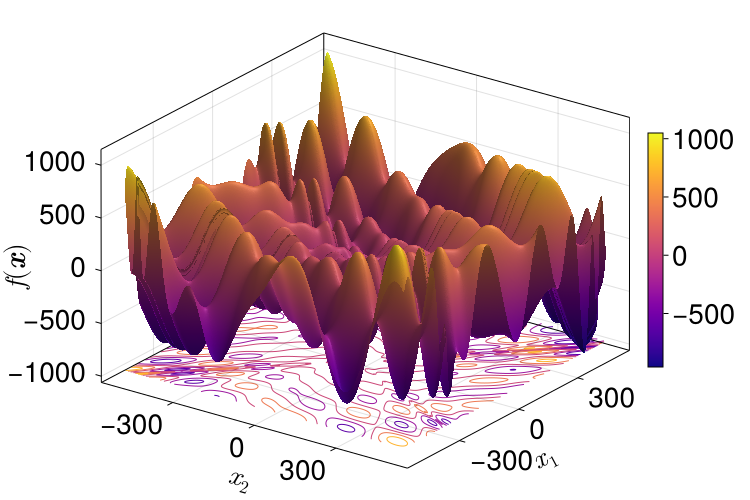}
        \caption{Egg Holder function}
    \end{subfigure}
    \begin{subfigure}[b]{0.33\textwidth}
        \includegraphics[width=\textwidth]{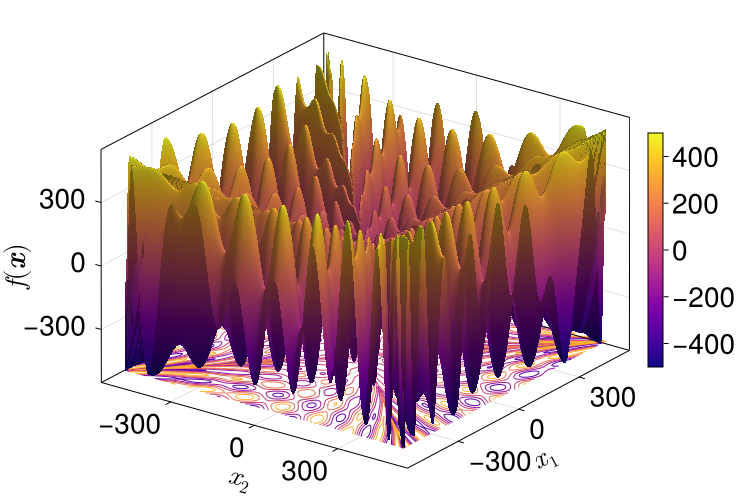}
        \caption{Rana's function}
    \end{subfigure}
    \caption{The landscapes of some 2D test functions in the continuous domain~\cite{vanaretCertifiedGlobalMinima2020}. The $z$-axis is used to plot the fitness, while the $x$- and $y$-axes are function parameters (shown here as $x_1$ and $x_2$).}
    \label{fig:fxs}
    
\end{figure}

Visualizing search landscapes of real-valued functions of low dimensionality ($\mathbb{R}^1, \mathbb{R}^2$ or $\mathbb{R}^3$) is straightforward. For higher dimensionalities, one must find a mapping from $\mathbb{R}^n$ to $\mathbb{R}^2$ or $\mathbb{R}^3$, and then plot the function to get a sense of the search space.\footnote{Finding a mapping from $\mathbb{R}^n$ to $\mathbb{R}^2$ or $\mathbb{R}^3$ is not trivial~\cite{massonVisualizingPseudoBooleanFunctions2025}.}
This mapping forms a continuous surface that can be analyzed with calculus tools, and our brains can wrap around it due to this continuity in the \textit{space}.
In this way, the features of the landscape (like peaks, valleys and funnels) are easily identified.
See for example Figure~\ref{fig:fxs} where different landscapes of 2D continuous test functions are plotted.

The situation is different when we deal with combinatorial landscapes since $\mathcal{X}$ is a discrete set.
What before was a smooth \textit{dune} in the continuous case, becomes a set of \textit{platforms} floating in mid-air, and the search points might not be able to be ordered---distance and continuity might not be well defined in the combinatorial case.
We discuss methods for visualizing combinatorial search landscapes in Section \ref{sec:vis}, paying special attention to those highlighting multimodality and their use of geoms and aesthetics.
For a more in-depth discussion of the mathematical implications of combinatorial landscapes, we refer the reader to the work of Reidys and Stadler~\cite{reidysCombinatorialLandscapes2002}.

\subsection{Multimodality in Landscape Analysis}
\label{sec:multimodality}

Multimodality, in the context of optimization, refers to the presence of multiple \textit{modes} in $\mathcal{X}$, i.e., the presence of multiple optima in the landscape.
When doing multimodal optimization, the goal is to \textit{extract the full set of optima and optimizers the problem possesses}~\cite{preussMultimodalOptimizationMeans2015}.
Realistically, in practice we look for multiple local optima for different reasons---to find different alternatives to a solution (to foster diversity~\cite{antipov_local_2024} or for solution robustness~\cite{dang_populations_2017}), to understand the structure of the problem and improve our search algorithms~\cite{mersmannExploratoryLandscapeAnalysis2011, ochoaLocalOptimaNetworks2014, thomsonEntropySearchTrajectories2024}, as well as a stepping stone to avoiding premature convergence to find the global optimum~\cite{liSeekingMultipleSolutions2017, sanchez-diaz_estimating_2024}.

Several \textit{everyday} activities of an AI practitioner involve multimodality. Consider the problem of feature selection in Machine Learning. It is a combinatorial optimization problem that is inherently multimodal (as several subsets of features can lead to the same performance of an ML model~\cite{liefoogheContrastingLandscapesFeature2024, sanchez-diazRegularizedFeatureSelection2024}).
Another multimodal combinatorial problem in the field of AI is hyper-parameter optimization: different combinations of hyper-parameter values can achieve the same classification results~\cite{mengshoel_understanding_2022, schneider_hpo_2022}.
Yet another example is the problem of neural network architecture search, where different architectures can achieve the same performance~\cite{lu_nsga-net_2019, stanley_designing_2019}.
These are only a handful of examples of multimodal optimization problems in AI, and they are all combinatorial in nature.
Therefore, a push for more studies on multimodality in optimization would be beneficial for both the evolutionary computation and the landscape analysis community.

\section{Related Work}
\label{sec:vis}

\subsection{Distance--Fitness Correlation}

\begin{figure}
    \centering
    \includegraphics[width=0.9\textwidth]{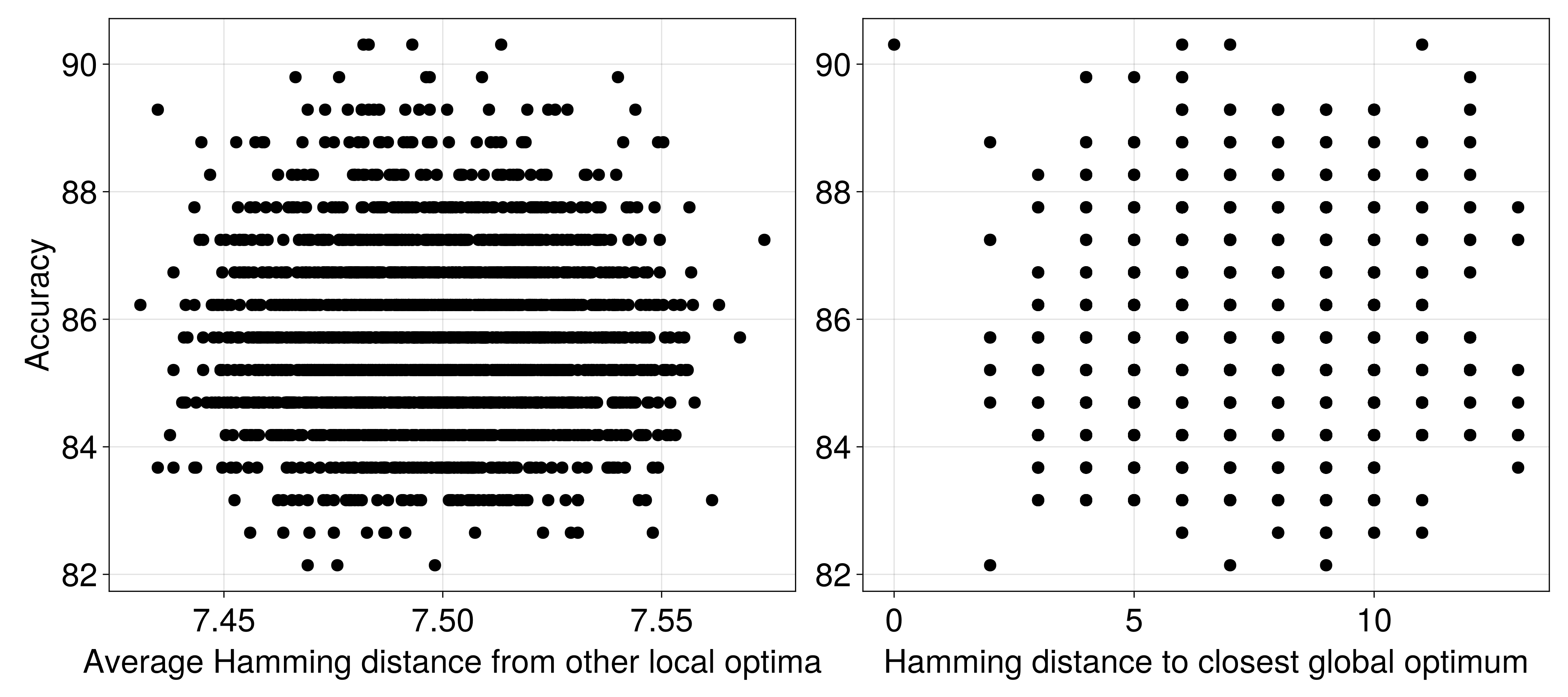}
    \caption{Analysis of 2511 local optima, on the feature selection problem of the Credit Approval~\cite{credit_approval_27} dataset using a decision tree classifier.
    The $y$-axis shows the quality of a solution (accuracy of classification) while the $x$-axis shows average Hamming distance between local optima (cf. left panel), and Hamming distance from each local optima to its closest global optimum (cf. right panel).
    }
    \label{fig:dist-corr}
    
\end{figure}

One of the key concepts in landscape analysis is the presence of a hypothetical \textit{big valley}: a region in the search space where multiple local optima occur at around the same distance from a global optimum.
%
%
The big valley is a common feature in multimodal landscapes, and it is often associated with the presence of a big central \textit{funnel}~\cite{doyeDoublefunnelEnergyLandscape1999}.
A useful visualization technique to identify funnel-like structures in search landscapes is to plot the distance between local optima and the global optimum (or the closest, in case there are multiple global optima).
See for example Figure~\ref{fig:dist-corr}, which highlights many local optima $\boldsymbol{b}^+$ at the same distance from the closest global optimum $\boldsymbol{b}^*$.

\begin{figure}[tb]
    \centering
    \includegraphics[width=0.95\textwidth]{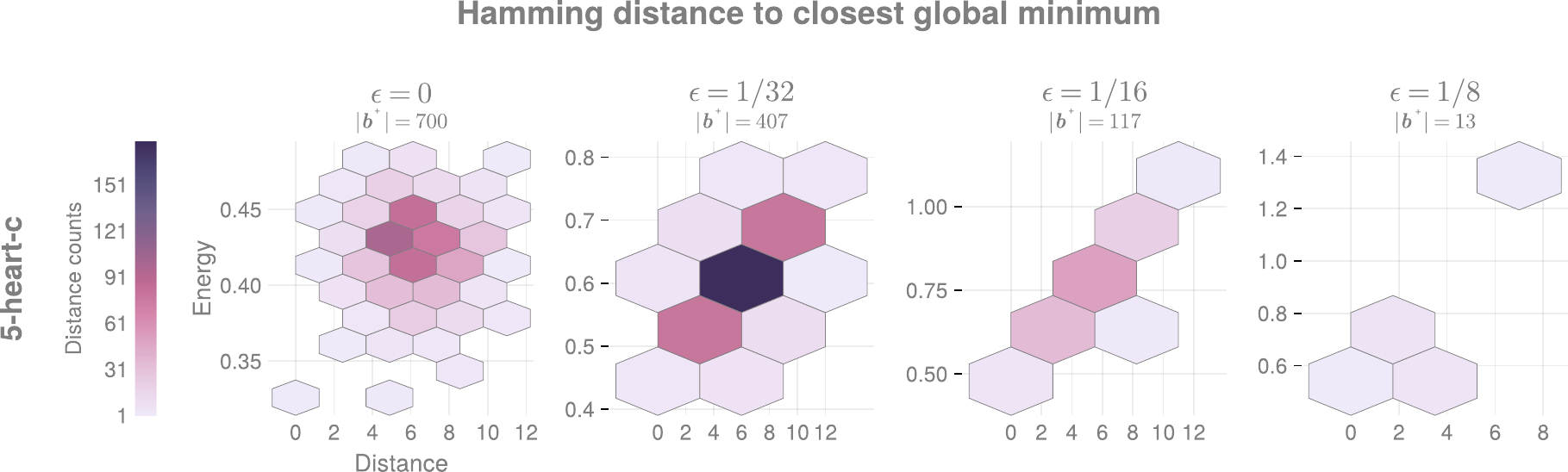}
    \caption{Hex-bin plot of distance correlation for the Heart Disease (Cleveland) dataset~\cite{heart_disease_45}, using a decision tree classifier under four different levels of regularization.
    Each bin aggregates different number of local optima, $\boldsymbol{b}^+$. A darker shade means a higher concentration of $\boldsymbol{b}^+$.}
    \label{fig:hexbin}
    
\end{figure}

Another approach is to aggregate the local optima in bins, and then reporting the size of each bin to get a sense of the distribution of local optima in the landscape. Figure~\ref{fig:hexbin} shows this approach, illustrating the landscape of a feature selection problem under different values of regularization.

\subsection{Local Optima Networks (LONs)}

LONs are the de-facto standard for highlighting the multimodal structure of a search landscape~\cite{liefoogheContrastingLandscapesFeature2024, ochoaLandscapeAnalysisOptimisation2024, ochoaStudyNKLandscapes2008, sanchez-diazRegularizedFeatureSelection2024}.
LONs can be described as graphs where nodes represent the local optima, and the edges represent paths between them.
An additional aesthetic element used in LONs is size, where bigger nodes represent bigger \textit{basins of attraction}. Color can be used to further convey the size of the basin (as shown in Figure~\ref{fig:lons}), but can also be used to communicate the fitness of each optimum~\cite{sanchez-diazRegularizedFeatureSelection2024}.

Different versions of LONs exist, like the Monotonic LON (MLON) where escape edges that end up in a worse local optima are removed, and the Compressed MLONs where plateaus are compressed into a single node~\cite{ochoaRecentAdvancesFitness2019}.
Additionally, LONs have been extended to multi-objective landscapes.
The Pareto Local Optimal Solutions Network (PLOS-net), as these LONs are called, are plotted along a vertical axis to represent the different ranks of the optima~\cite{liefoogheParetoLocalOptimal2018}.
As with LONs, PLOS-nets can also be compressed to remove plateaus.


Both LONs and PLOS-nets give a visual overview of the multimodality of a landscape as well as its neighborhood or connectedness.
The main focus on these visualizations is to highlight the size of the basins of attraction, but the number and the distribution of the optima cannot be easily inferred from these plots.

\subsection{Hinged Bitstring Maps (HBMs)}

HBMs were introduced for visualizing pseudo-Boolean landscapes in PPSN 2024~\cite{sanchez-diazRegularizedFeatureSelection2024}.
HBMs plot the entire search space $\mathcal{X}$ by slicing the bitstring $\boldsymbol{b}$ into two halves, and using the first half (converted to its decimal representation) as the $x$-axis.
The $y$-axis is assigned to the decimal representation of the second half.
Each solution is then plotted as a dot at the $(x, y)$ coordinate, colored by its fitness value.
Optima can additionally be highlighted by using a colored outline.
We show an example of an HBM in Figure~\ref{fig:hbm}.

HBMs are designed to look at $\mathcal{X}$ in its entirety to highlight multimodality.
However, depicting the neighborhood is difficult, as HBMs can become cluttered for large values of $n$ since $|\mathcal{X}| = |\mathbb{B}^n| = 2^n$.
Nevertheless, one of the advantages of looking at the whole search space is that local optima can be `counted' and patterns about the distribution of optima can be analyzed.


\subsection{Sequence Index Plots}

As opposed to bitstrings, sequences do not have an inherent order. As such, visualizing sequences usually requires full enumeration.
Combinations can be represented as a path in a fully connected graph, and sequences can be represented as a path in a directed graph in a similar manner, but this approach becomes infeasible for large cardinalities of $\mathcal{X}$, and so a sequence index plot might be more suitable~\cite{fasangVisualizingSequencesSocial2014, gleicherConsiderationsVisualizingComparison2018}.
A sequence index plot uses the $x$-axis to represent the order of the sequence, and the $y$-axis to stack the different sampled sequences (or the \textit{states}, in optimization).
This type of visualization is useful when the number of sequences is small.
However, it can also become cluttered when the domain for each position in the sequence is large enough.

Alternative visualizations for sequences exist, but these often include fully-built graphical user interfaces in specialized software, where additional information can be provided via \textit{tooltips} or other interactive elements~\cite{gleicherConsiderationsVisualizingComparison2018, vanderlindenSurveyVisualizationTechniques2023}.
Among static graphics, LONs tend to be more suitable to represent search landscapes of sequences. HBMs could also be used if the number of sequences to be visualized is small, provided that the sequences can be enumerated.

\subsection{Search Trajectory Networks (STNs)}

A search trajectory network (STN) is a special case of LON. It is a directed graph that represents the search process of multiple optimization algorithms \cite{ochoaSearchTrajectoryNetworks2021}.
As the name implies, the network is made up of different \textit{trajectories}, which are paths in $\mathcal{X}$ that represent the sequence of states $\boldsymbol{b}$ visited by each algorithm.
Several algorithms can be represented in the same STN, and the edges can be weighted by the number of times a transition between two states has been made.

\begin{figure}
    \centering
    \includegraphics[width=0.55\textwidth]{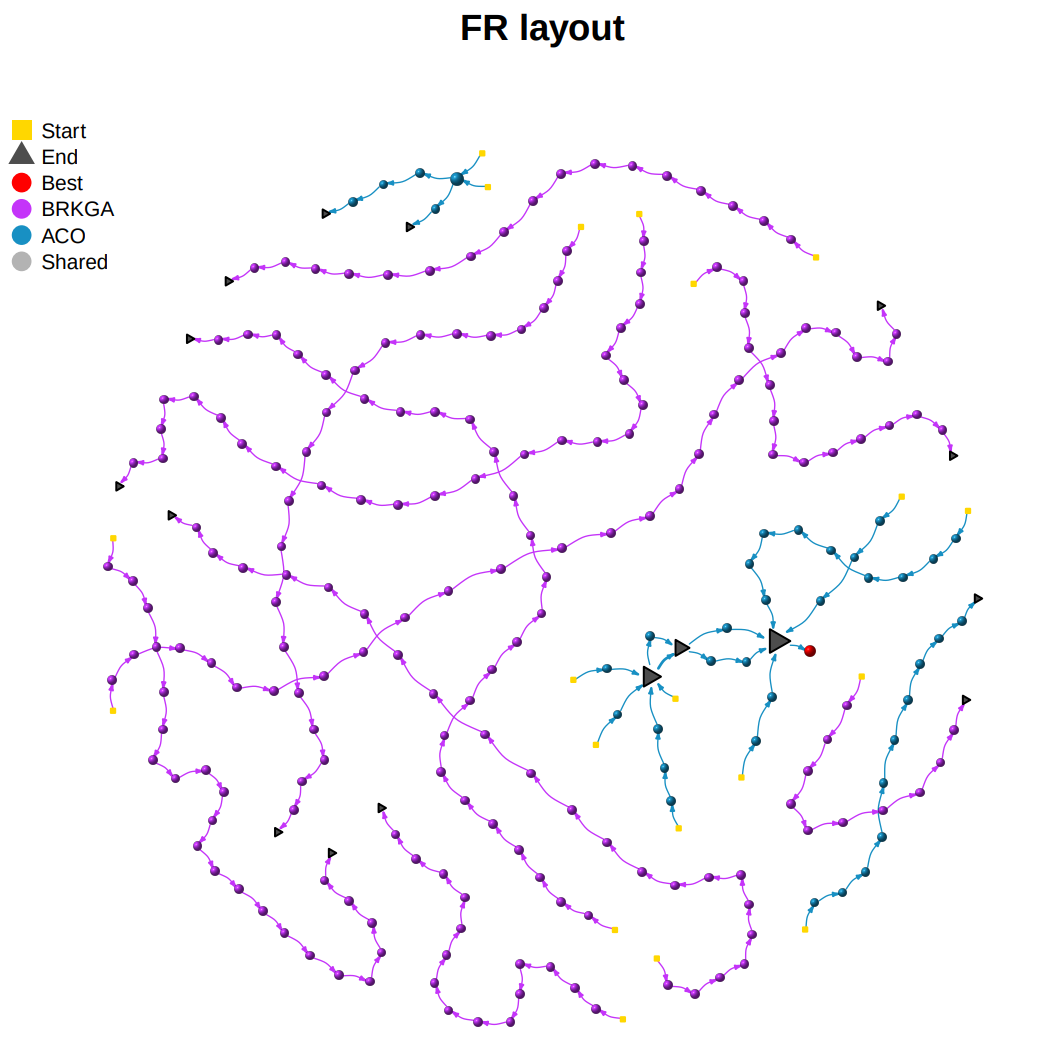}
    \caption{An STN comparing two algorithms: Biased Random-Key Genetic Algorithm and Ant Colony Optimization using the discrete example problem from Ochoa et al.~\cite{ochoaSearchTrajectoryNetworks2021}. The plot was generated using the STNs online tool.}
    \label{fig:stn}
    
\end{figure}

STNs can be used to compare different optimization algorithms in a single visualization, using different \textit{geoms} and aesthetics to represent the different heuristics.
Additionally, notable nodes are usually highlighted---like the best solution found by each algorithm, its start and end states, as well as ``shared'' states that were visited in different runs.
As is the case with LONs, STNs can be compressed depending on the granularity used to represent the states, and additional metrics can be calculated due to their graph structure.
An example of an STN is shown in Figure~\ref{fig:stn}, generated by the STN Analytics online tool using the discrete example problem.\footnote{See \url{https://www.stn-analytics.com/}.}

Multimodality can be inferred from STNS by looking at the shared nodes.
However, the notions of distance and neighborhoods between the optima are not as clear as with LONs, since these graphics were not specifically designed to highlight such features.

\subsection{Violation Landscapes (VL)}

An important feature of real-world optimization problems is the presence of constraints, and they occur in both continuous and combinatorial landscapes.
When $\mathcal{X}$ contains solutions that are not feasible, we say that the solution is in \textit{violation}.
In a constrained search space, optimization algorithms must navigate the landscape to find feasible solutions, and it can be challenging when the constraints are complex or when the feasible region is small.
It is, therefore, important to visualize the violation landscape (VL), which is the subset of $\mathcal{X}$ that is feasible.

\begin{figure}[tb]
    \centering
    \includegraphics[width=0.95\textwidth]{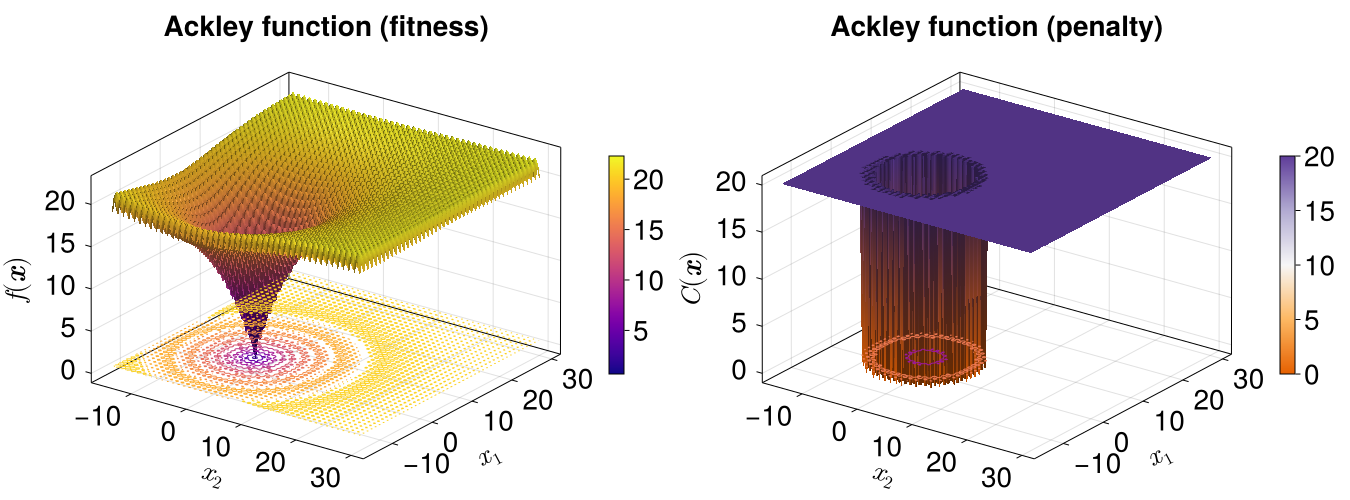}
    \caption{Fitness and violation landscapes of a 2D Constrained Ackley Function. The constraint is handled as a penalty for all solutions $\boldsymbol{b}$ with $f(\boldsymbol{b})>15$, creating two regions in $\mathcal{X}$.}
    \label{fig:constraints}
    
\end{figure}

One way to visualize the VL is to plot the fitness of the solutions in $\mathcal{X}$, but only showing the feasible solutions.
Another (and perhaps the most common) approach is to use color in the plot to differentiate between the feasible and infeasible solutions.
This seems more intuitive when the $\mathcal{X}$ is continuous~\cite{malanCharacterisingConstrainedContinuous2015, mallipeddiProblemDefinitionsEvaluation2010}. Figure~\ref{fig:constraints} shows an example of a 2D continuous functions with a constraint, the Ackley function.

As shown in Figure~\ref{fig:constraints}, VL visualizations use color to represent the feasibility of a solution which has a location determined by its $x$ and $y$ coordinates in the plot.
To have a better overview of the problem instance, both fitness and violation landscapes are usually presented next to one another~\cite{malanCharacterisingConstrainedContinuous2015, mallipeddiProblemDefinitionsEvaluation2010}.
However, using color to show infeasibility is not the only way to plot a VL.
In fact, the VL itself depends on the definition of \textit{constraint violation}, and in turn, different visualizations can be generated depending on the definitions used~\cite{yasudaAnalyzingViolationLandscapes2024}.

Combinatorial landscapes with constraints are more difficult to visualize, since the same issue as with unconstrained landscapes persists, but now with the additional requirement of adding solution feasibility to the plot.
However, the same principles can be applied: color can be used to represent the feasibility of a solution, and the $x$ and $y$ axes can be used to represent the solution itself (as in Figure~\ref{fig:constraints}).

\section{Combining Visualization Techniques}
\label{sec:combining}

\begin{figure}[htb]
    \centering
    \includegraphics[width=0.9\textwidth]{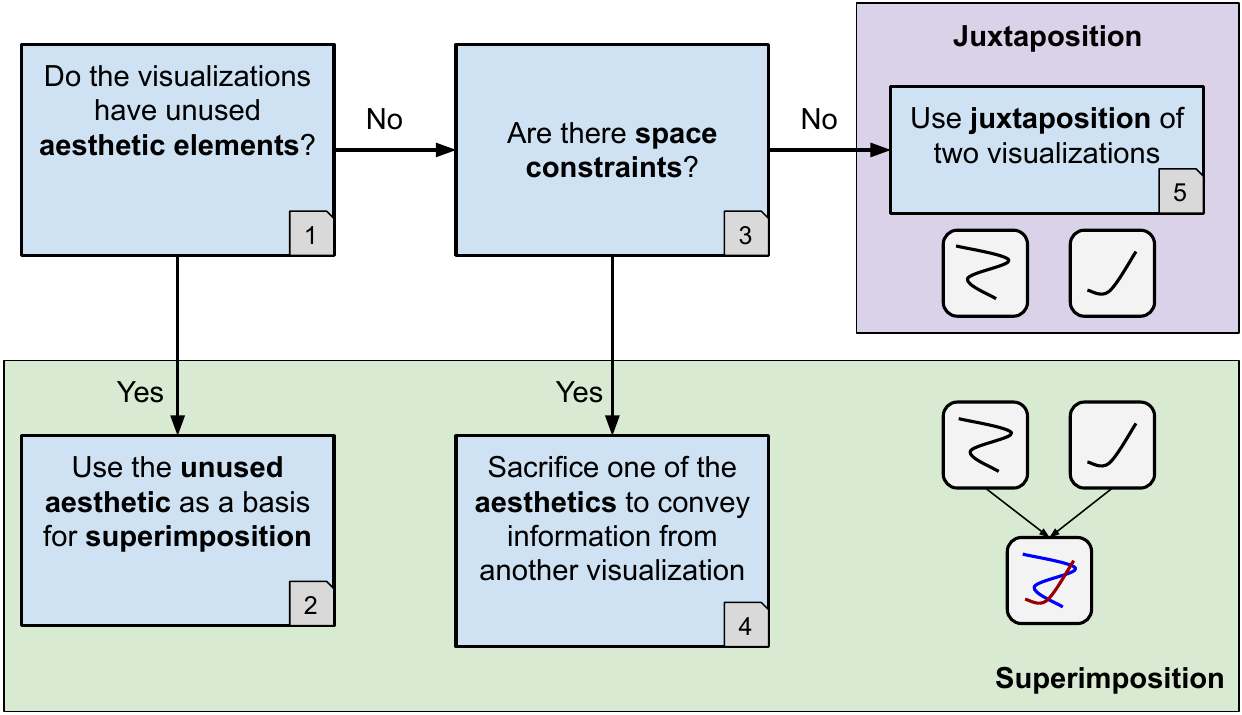}
    \caption{A simple process for combining visualizations. Superimposition may require additional data transformations, while juxtaposition demands more space.}
    \label{fig:vizmerge}
    
\end{figure}

Different visualization techniques are not mutually exclusive.
In fact, combining them can provide a more comprehensive view of a combinatorial landscape, since different visualizations highlight different aspects of the search space.
Several ways of merging visualizations exist~\cite{gleicherConsiderationsVisualizingComparison2018, nobreStateArtVisualizing2019}, but the most common are juxtaposition, superimposition, overloading and nesting~\cite{javedExploringDesignSpace2012}.
In this work we focus on the first two as a basis for combining visualization techniques, and present this process in Figure~\ref{fig:vizmerge}.

\subsection{Juxtaposition}

Placing different visualizations next to one another is the simplest way to have different views of the same landscape. However, it requires an efficient relational linking and sufficient space to have a comfortable layout~\cite{javedExploringDesignSpace2012}.
As an example of juxtaposition, consider aligning two binary strings (one on top of another).
This view allows for immediate recognition of bits that are different at a given position, and it can be useful to identify, for example, redundant features in feature selection~\cite{mostert_feature_2021, sanchez-diazRegularizedFeatureSelection2024}.

Most of the figures in this work (Figures~\ref{fig:dist-corr}, \ref{fig:hexbin}, \ref{fig:constraints}, \ref{fig:lons} and \ref{fig:hbm}) are examples of juxtaposition. We now discuss two case studies.

\subsubsection{Juxtaposition of LONs: Zooming in on $\mathcal{L}$}

\begin{figure}[tb]
    \centering
    \begin{subfigure}[b]{0.47\textwidth}
        \includegraphics[width=\textwidth]{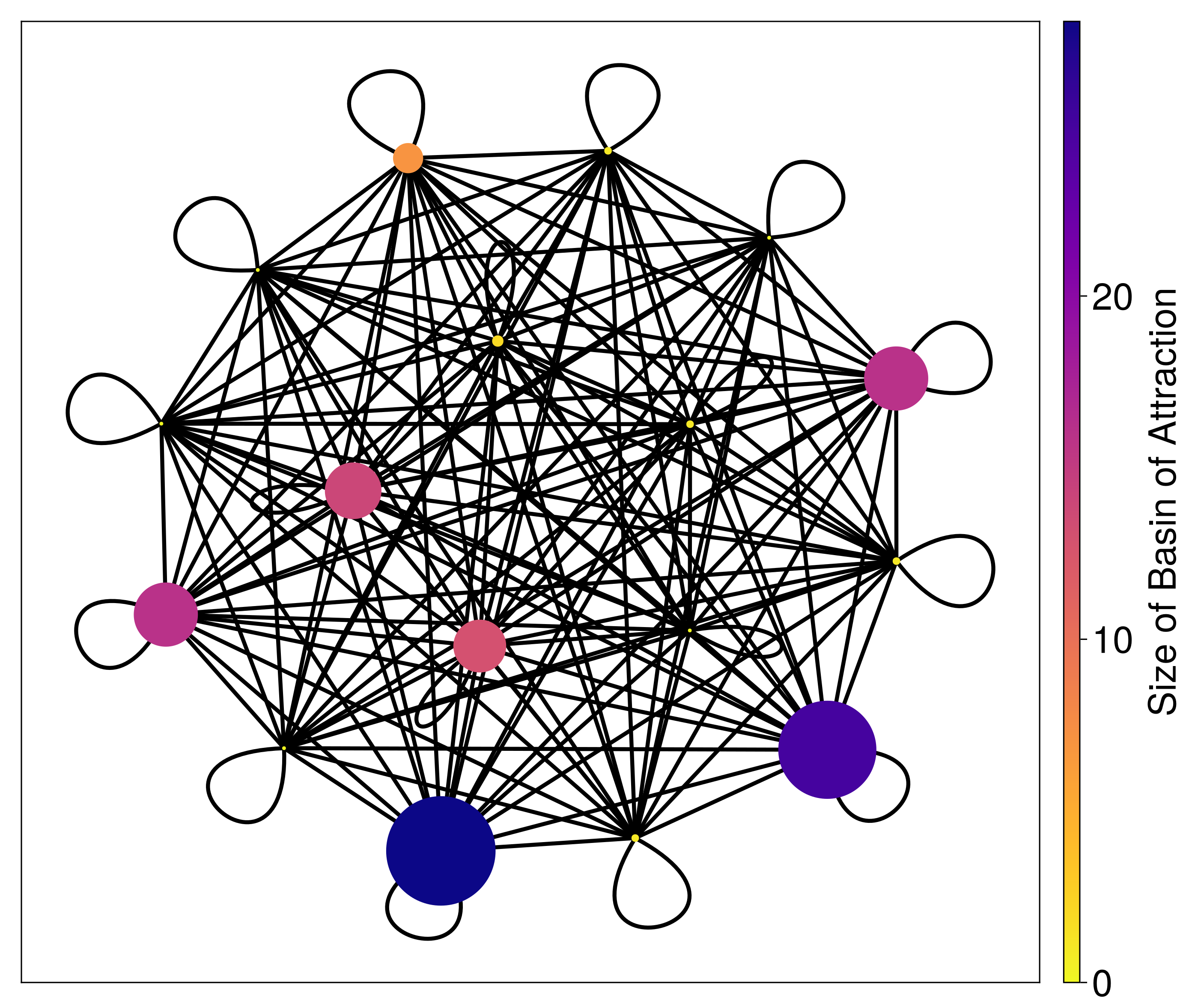}
        \caption{A LON with all basin transition edges}
    \end{subfigure}%
    \begin{subfigure}[b]{0.47\textwidth}
        \includegraphics[width=\textwidth]{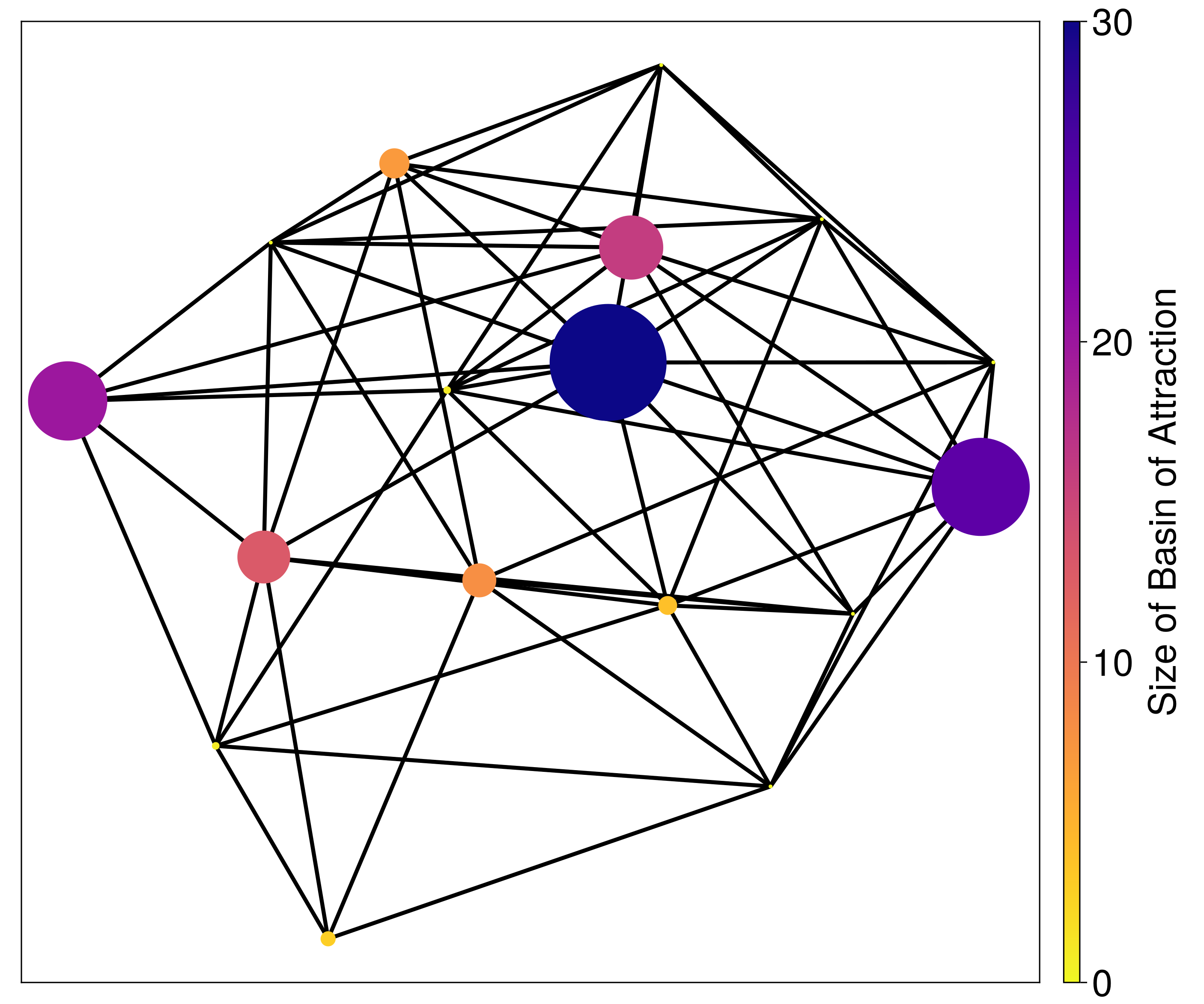}
        \caption{A LON with escape edges with $D=3$}
        \label{fig:escape}
    \end{subfigure}%
    \caption{LON-LON: Juxtaposition of two LONs, representing the feature selection problem on the E-coli dataset~\cite{ecoli_39}, using a decision tree classifier. In \ref{fig:escape}, the escape edges are only kept if their Hamming distance $d_H \leq D$.}
    \label{fig:lons}
    
\end{figure}

A first obvious case of study is the juxtaposition of two different views of the same object. In Figure~\ref{fig:lons}, we present two LON-based views of the same landscape: a feature selection problem sampled with different levels of detail.
On the left panel of Figure \ref{fig:lons}, a LON with all basin transitions is presented, while the right panel shows a LON with only escape edges of size 3 or less.
The former presents an overview of the whole landscape (and the connectedness between all the 17 local optima), while the latter presents a detailed view of only those local optima that can be reached in three or fewer bit flips from the other optima.
In this case, juxtaposition allows us to switch the view and \textit{zoom-in} on the details of the important elements of the landscape.

\subsubsection{Juxtaposition of HBMs: Transformations on $f$}

\begin{figure}[tb]
    \centering
    \includegraphics[width=\textwidth]{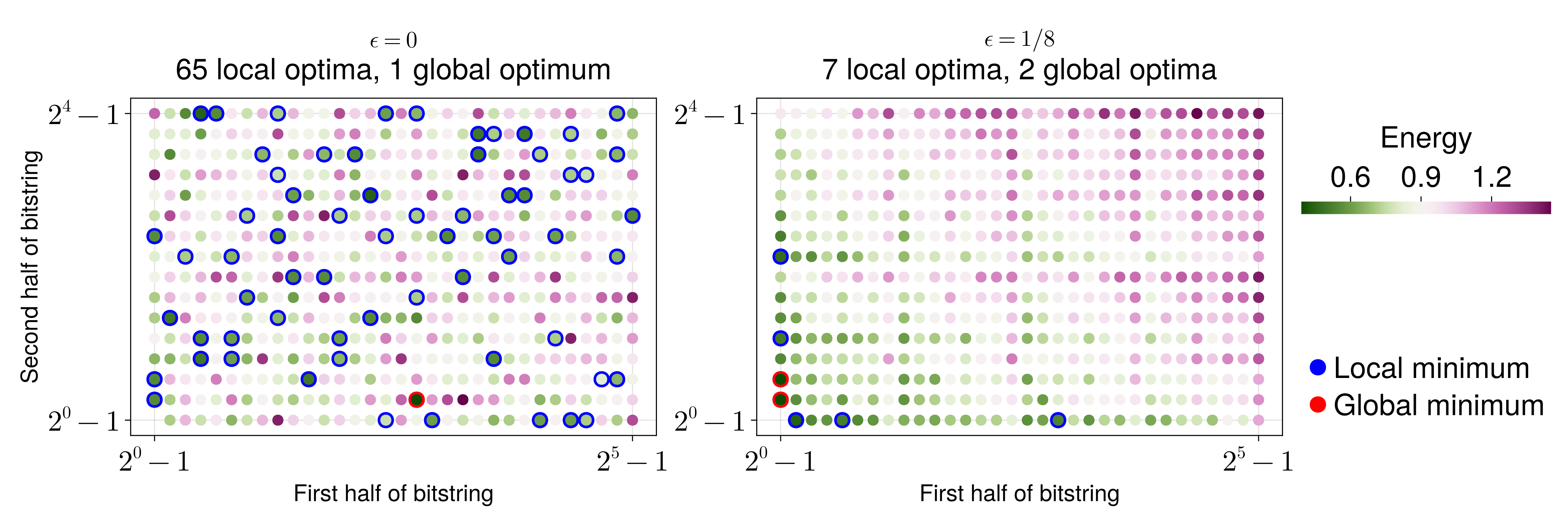}
    \caption{HBM-HBM: Juxtaposition of two HBMs for the Glass Identification dataset~\cite{glass_identification_42} with different values of regularization. The local and global optima are highlighted with blue and red outlines, respectively.}
    \label{fig:hbm}
    
\end{figure}

Since the HBM was designed to look at the whole search space, transformations of the fitness function can be easily visualized by placing two HBMs side by side as long as the transformations on $f$ have the same domain ($\mathcal{X}$).
In Figure~\ref{fig:hbm} we show the landscape of a feature selection problem under two different degrees of regularization.
Here we used a colored outline to highlight local and global optima, and can compare how the number (and distribution) of these optima changes when the landscape is transformed.
In this way, juxtaposition helps us identify that the $\epsilon=1/8$ regularization (cf. right panel in Figure~\ref{fig:hbm}) reduces the number of optima that originally existed when $\epsilon=0$ (left panel in Figure~\ref{fig:hbm}).

\subsection{Superimposition}

A superimposed view can be obtained when a visualization is overlaid on top of another, resulting in a new view that combines elements of both.
These views are usually employed to highlight \textit{spatial} relations between the original visualizations since there is a one-to-one correspondence between their spatial linking~\cite{javedExploringDesignSpace2012}.
For instance, showing several routes in a Vehicle Routing or a Traveling Salesman Problem on top of a map allows for an easy comparison between them.

Superimposition of different kinds of landscape visualizations can be done in a similar fashion if there are \textit{free} or \textit{unused} aesthetics on at least one of them.
Table~\ref{tab:combo} shows a summary of the different aesthetics and geoms used by the three main visualization techniques studied in this work---LONs, HBMs, and STNs.
The aesthetic attributes that are not in use in one visualization can be used to represent information from the other, as long as there exists a mapping between them.
A transformation might be needed to \textit{fit} one of the aesthetics on top of another, but it can lead to more informative designs, even when there is no isometric projection between their axes~\cite{massonVisualizingPseudoBooleanFunctions2025}.
We now discuss a case study of superimposition, describing how to combine a LON and an HBM.

\begin{table}[tb]
\centering
\caption{Different aesthetic and geoms used by the three main visualization techniques studied. Attributes marked with N/A, also called unused aesthetic elements, are not used in the visualization and are thus free to be used for other purposes, as is the case with the LON+HBM (superimposition).}
\label{tab:combo}
\resizebox{\textwidth}{!}{%
\begin{tabular}{@{}lllllll@{}}
\toprule
\multicolumn{1}{c}{\multirow{2}{*}{\textbf{Plot type}}} & \multicolumn{2}{c}{\textbf{Geoms}} & \multicolumn{4}{c}{\textbf{Aesthetics}} \\ \cmidrule(l){2-7} 
\multicolumn{1}{c}{} & \multicolumn{1}{c}{\textbf{Primary}} & \multicolumn{1}{c}{\textbf{Secondary}} & \multicolumn{1}{c}{\textbf{Color}} & \multicolumn{1}{c}{\textbf{Size}} & \multicolumn{1}{c}{\textbf{Position}} & \multicolumn{1}{c}{\textbf{Visibility}} \\ \midrule
LON & Circle & Lines & Basin of attraction & Basin of attraction & N/A & $\mathcal{L} \subset \mathcal{X}$ \\
HBM & Circle & Rings & Fitness & N/A & $\boldsymbol{b}$ & $\mathcal{X}$ \\
STN & Circle & Arrows & Algorithm & Frequency & N/A & Explored space \\
 & Triangle &  &  &  &  &  \\
 & Square &  &  &  &  &  \\ \midrule
\multicolumn{7}{c}{\textbf{Combined visualizations}} \\ \midrule
LON-LON & Circle & Lines & Basin of attraction & Basin of attraction & N/A & $\mathcal{L} \subset \mathcal{X}$ \\
HBM-HBM & Circle & Rings & Fitness & N/A & $\boldsymbol{b}$ & $f \circ \mathcal{X}, f' \circ \mathcal{X}$ \\
LON+HBM & Circle & Lines & Basin of attraction & Basin of attraction & $\boldsymbol{b}$ & $\mathcal{X}$ \\ \bottomrule
\end{tabular}
}

\end{table}

\subsubsection{Superimposition of LON+HBM: Identification of $\mathcal{L}$}

\begin{figure}[tb]
    \centering
    \includegraphics[width=0.9\textwidth]{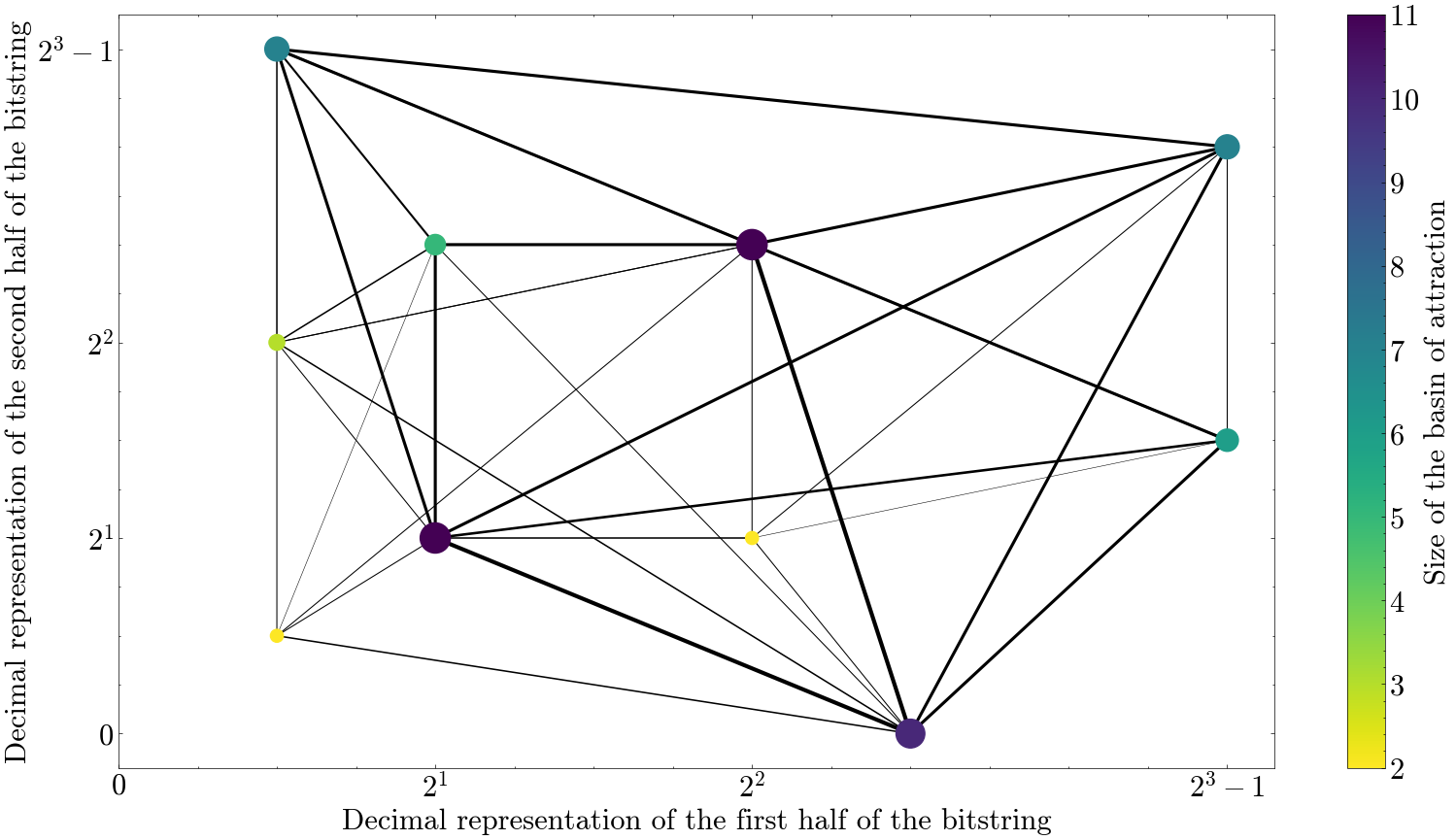}
    \caption{LON+HBM: a LON using the HBM coordinate system to visualize the landscape of $f(\boldsymbol{b})=\sin\left(2\,\mathrm{Dec}(\boldsymbol{b})\right), \, \forall \boldsymbol{b} \in \mathbb{B}^6$.
    The color is used to represent the size of the basin of attraction. Figure reproduced with permission from Masson et al.~\cite{massonVisualizingPseudoBooleanFunctions2025}.}
    \label{fig:lonhbm}
    
\end{figure}

A LON+HBM is a good example of superimposition as it gives a glimpse into the multimodality of the landscape and the distribution of the optima.
As an example, see Figure~\ref{fig:lonhbm}, where a LON is plotted on top of an HBM depicting the landscape of toy problem: $f(\boldsymbol{b})=\sin\left(2\,\mathrm{Dec}(\boldsymbol{b})\right), \, \forall \boldsymbol{b} \in \mathbb{B}^6$.
The used aesthetics are the same as in the standalone plots, but we use the unused attributes of one to merge into the other: LONs do not have a natural mapping to the $x$- and $y$-axes, but using the HBM coordinate system, we can place the nodes in space using the same hinged-mapping.
Thus, identification of the location of a local optimum is possible, as well as working out which optimizer $\boldsymbol{b}^+$ corresponds to a specific basin of attraction.

\subsection{No Free Lunch in Landscape Visualization}

Even when there are \textit{free} attributes and a compromise can be made to merge two visualizations, some combinations can become difficult to read.
For example, using the unused aesthetic attribute of STNs, we can plot the sequence of visited solutions on top of an HBM.
However, the mapping between the sequence of visited states and the HBM would be difficult to establish, as the $x$- and $y$-axes of the HBM are not directly related to the sequence of the search.
A possible workaround is to use text labels inside each node to represent the sequence of a specific algorithm, and use color to show different algorithms.
The fitness would need to be taken out of the plot, with the opportunity to compensate by plotting another HBM next to it.
It should not take long to realize that this visualization would quickly become difficult to read, as following different number sequences using different colors for the foreground text and the background shape requires a lot of focus.
Nevertheless, it is a good example to understand the limitations of the different visualization techniques.

On the other hand, combining an HBM and a VL is trivial: using color to plot the feasibility of a solution instead of the fitness.
This works well because the \textit{position} aesthetic in both visualization techniques is used to represent the same feature of the landscape, which is the location of the solution itself.
Then, to convey more information, the feasibility HBM can be \textit{juxtaposed} next to an HBM with fitness, and the distribution of both feasible solutions and the optima over the landscape can be observed.

\section{Conclusions and Recommendations}
\label{sec:conc}

In this work we discussed the importance of visualizing multimodality in combinatorial search landscapes.
We have shown that different visualization techniques have their own strengths and limitations, and provided a simple way to combine them as their composition can generate more comprehensive views of the landscape.
With the Grammar of Graphics, we have shown that the different aesthetics and geoms can be employed to represent different features of the landscape, and that the unused aesthetic attributes of one plot can extend the expressive power of another.

For future work, the use of another aesthetics might be worth visiting. In particular, the time aesthetic---animated visualizations can provide a detailed view of the landscape in dynamic or noisy scenarios.
Considering other landscape visualization methods (e.g. attractor networks~\cite{thomson_stalling_2024}) is another interesting path for future work.

\section*{Acknowledgments}
We would like to acknowledge the financial support from Project no. 311284 of the Research Council of Norway, as well as access to computing resources from the Department of Computer Science of the Norwegian University of Science and Technology.

\section*{Declaration on Generative AI}
The authors have not employed any Generative AI tools.

\printbibliography

@article{boeseNewAdaptiveMultistart1994,
  title = {A New Adaptive Multi-Start Technique for Combinatorial Global Optimizations},
  author = {Boese, Kenneth D. and Kahng, Andrew B. and Muddu, Sudhakar},
  year = {1994},
  month = sep,
  journal = {Operations Research Letters},
  volume = {16},
  number = {2},
  pages = {101--113},
  issn = {0167-6377},
  doi = {10.1016/0167-6377(94)90065-5},
  urldate = {2024-11-25},
}

@article{doyeDoublefunnelEnergyLandscape1999,
  title = {The Double-Funnel Energy Landscape of the 38-Atom {{Lennard-Jones}} Cluster},
  author = {Doye, Jonathan P. K. and Miller, Mark A. and Wales, David J.},
  year = {1999},
  month = apr,
  journal = {The Journal of Chemical Physics},
  volume = {110},
  number = {14},
  pages = {6896--6906},
  issn = {0021-9606},
  doi = {10.1063/1.478595},
  urldate = {2024-11-28},
}

@article{fasangVisualizingSequencesSocial2014,
  title = {Visualizing {{Sequences}} in the {{Social Sciences}}: {{Relative Frequency Sequence Plots}}},
  shorttitle = {Visualizing {{Sequences}} in the {{Social Sciences}}},
  author = {Fasang, Anette Eva and Liao, Tim Futing},
  year = {2014},
  journal = {Sociological Methods \& Research},
  volume = {43},
  number = {4},
  pages = {643--676},
  publisher = {Thousand Oaks, CA: Sage},
  issn = {1552-8294},
  doi = {10.1177/0049124113506563},
  urldate = {2024-11-23},
  copyright = {http://www.econstor.eu/dspace/Nutzungsbedingungen},
  langid = {english},
}

@inproceedings{liefoogheContrastingLandscapesFeature2024,
  title = {Contrasting the~{{Landscapes}} of~{{Feature Selection Under Different Machine Learning Models}}},
  booktitle = {Parallel {{Problem Solving}} from {{Nature}} -- {{PPSN XVIII}}},
  author = {Liefooghe, Arnaud and Tanabe, Ryoji and Verel, S{\'e}bastien},
  editor = {Affenzeller, Michael and Winkler, Stephan M. and Kononova, Anna V. and Trautmann, Heike and Tu{\v s}ar, Tea and Machado, Penousal and B{\"a}ck, Thomas},
  year = {2024},
  pages = {360--376},
  publisher = {Springer Nature Switzerland},
  address = {Cham},
  doi = {10.1007/978-3-031-70055-2_22},
  isbn = {978-3-031-70055-2},
  langid = {english},
}

@inproceedings{liefoogheParetoLocalOptimal2018,
  title = {On {{Pareto Local Optimal Solutions Networks}}},
  booktitle = {Parallel {{Problem Solving}} from {{Nature}} -- {{PPSN XV}}},
  author = {Liefooghe, Arnaud and Derbel, Bilel and Verel, S{\'e}bastien and {L{\'o}pez-Ib{\'a}{\~n}ez}, Manuel and Aguirre, Hern{\'a}n and Tanaka, Kiyoshi},
  editor = {Auger, Anne and Fonseca, Carlos M. and Louren{\c c}o, Nuno and Machado, Penousal and Paquete, Lu{\'i}s and Whitley, Darrell},
  year = {2018},
  pages = {232--244},
  publisher = {Springer International Publishing},
  address = {Cham},
  doi = {10.1007/978-3-319-99259-4_19},
  isbn = {978-3-319-99259-4},
  langid = {english},
}

@article{liSeekingMultipleSolutions2017,
  title = {Seeking {{Multiple Solutions}}: {{An Updated Survey}} on {{Niching Methods}} and {{Their Applications}}},
  shorttitle = {Seeking {{Multiple Solutions}}},
  author = {Li, Xiaodong and Epitropakis, Michael G. and Deb, Kalyanmoy and Engelbrecht, Andries},
  year = {2017},
  month = aug,
  journal = {IEEE Transactions on Evolutionary Computation},
  volume = {21},
  number = {4},
  pages = {518--538},
  issn = {1941-0026},
  doi = {10.1109/TEVC.2016.2638437},
  urldate = {2023-09-25},
}

@misc{MakieOrgAlgebraOfGraphicsjl2024,
  title = {{{MakieOrg}}/{{AlgebraOfGraphics}}.Jl},
  year = {2024},
  month = nov,
  urldate = {2024-11-29},
  copyright = {MIT},
  howpublished = {MakieOrg}
}

@inproceedings{malanCharacterisingConstrainedContinuous2015,
  title = {Characterising Constrained Continuous Optimisation Problems},
  booktitle = {2015 {{IEEE Congress}} on {{Evolutionary Computation}} ({{CEC}})},
  author = {Malan, Katherine M. and Oberholzer, Johannes F. and Engelbrecht, Andries P.},
  year = {2015},
  month = may,
  pages = {1351--1358},
  issn = {1941-0026},
  doi = {10.1109/CEC.2015.7257045},
  urldate = {2024-11-29},
}

@inproceedings{malanRecentAdvancesLandscape2021,
  title = {Recent Advances in Landscape Analysis for Optimisation and Learning},
  booktitle = {Proceedings of the {{Genetic}} and {{Evolutionary Computation Conference Companion}}},
  author = {Malan, Katherine and Ochoa, Gabriela},
  year = {2021},
  month = jul,
  series = {{{GECCO}} '21},
  pages = {899--917},
  publisher = {Association for Computing Machinery},
  address = {New York, NY, USA},
  doi = {10.1145/3449726.3461396},
  urldate = {2024-01-27},
  isbn = {978-1-4503-8351-6},
}

@inproceedings{mallipeddiProblemDefinitionsEvaluation2010,
  title = {Problem {{Definitions}} and {{Evaluation Criteria}} for the {{CEC}} 2010 {{Competition}} on {{Constrained Real- Parameter Optimization}}},
  author = {Mallipeddi, R. and Suganthan, P.},
  year = {2010},
  urldate = {2024-12-01}
}

@inproceedings{massonVisualizingPseudoBooleanFunctions2025,
  title = {Visualizing {{Pseudo-Boolean Functions}}: {{Feature Selection}} and~{{Regularization}} for~{{Machine Learning}}},
  shorttitle = {Visualizing {{Pseudo-Boolean Functions}}},
  booktitle = {Evolutionary {{Computation}} in {{Combinatorial Optimization}}},
  author = {Masson, Corentin and {S{\'a}nchez-D{\'i}az}, Xavier F. C. and Mengshoel, Ole Jakob},
  editor = {Krejca, Martin S. and Wagner, Markus},
  year = {2025},
  pages = {150--166},
  publisher = {Springer Nature Switzerland},
  address = {Cham},
  doi = {10.1007/978-3-031-86849-8_10},
  isbn = {978-3-031-86849-8},
  langid = {english},
}

@inproceedings{mersmannExploratoryLandscapeAnalysis2011,
  title = {Exploratory Landscape Analysis},
  booktitle = {Proceedings of the 13th Annual Conference on {{Genetic}} and Evolutionary Computation},
  author = {Mersmann, Olaf and Bischl, Bernd and Trautmann, Heike and Preuss, Mike and Weihs, Claus and Rudolph, G{\"u}nter},
  year = {2011},
  month = jul,
  series = {{{GECCO}} '11},
  pages = {829--836},
  publisher = {Association for Computing Machinery},
  address = {New York, NY, USA},
  doi = {10.1145/2001576.2001690},
  urldate = {2024-01-22},
  isbn = {978-1-4503-0557-0},
}

@inproceedings{ochoaLandscapeAnalysisOptimisation2024,
  title = {Landscape {{Analysis}} of {{Optimisation Problems}} and {{Algorithms}}},
  booktitle = {Proceedings of the {{Genetic}} and {{Evolutionary Computation Conference Companion}}},
  author = {Ochoa, Gabriela and Malan, Katherine},
  year = {2024},
  month = aug,
  series = {{{GECCO}} '24 {{Companion}}},
  pages = {984--1004},
  publisher = {Association for Computing Machinery},
  address = {New York, NY, USA},
  doi = {10.1145/3638530.3648421},
  urldate = {2024-11-24},
  isbn = {979-8-4007-0495-6},
}

@incollection{ochoaLocalOptimaNetworks2014,
  title = {Local {{Optima Networks}}: {{A New Model}} of {{Combinatorial Fitness Landscapes}}},
  shorttitle = {Local {{Optima Networks}}},
  booktitle = {Recent {{Advances}} in the {{Theory}} and {{Application}} of {{Fitness Landscapes}}},
  author = {Ochoa, Gabriela and Verel, S{\'e}bastien and Daolio, Fabio and Tomassini, Marco},
  editor = {Richter, Hendrik and Engelbrecht, Andries},
  year = {2014},
  pages = {233--262},
  publisher = {Springer},
  address = {Berlin, Heidelberg},
  doi = {10.1007/978-3-642-41888-4_9},
  urldate = {2024-11-29},
  isbn = {978-3-642-41888-4},
  langid = {english},
}

@inproceedings{ochoaRecentAdvancesFitness2019,
  title = {Recent Advances in Fitness Landscape Analysis},
  booktitle = {Proceedings of the {{Genetic}} and {{Evolutionary Computation Conference Companion}}},
  author = {Ochoa, Gabriela and Malan, Katherine},
  year = {2019},
  month = jul,
  series = {{{GECCO}} '19},
  pages = {1077--1094},
  publisher = {Association for Computing Machinery},
  address = {New York, NY, USA},
  doi = {10.1145/3319619.3323383},
  urldate = {2024-01-29},
  isbn = {978-1-4503-6748-6},
}

@article{ochoaSearchTrajectoryNetworks2021,
  title = {Search Trajectory Networks: {{A}} Tool for Analysing and Visualising the Behaviour of Metaheuristics},
  shorttitle = {Search Trajectory Networks},
  author = {Ochoa, Gabriela and Malan, Katherine M. and Blum, Christian},
  year = {2021},
  month = sep,
  journal = {Applied Soft Computing},
  volume = {109},
  pages = {107492},
  issn = {1568-4946},
  doi = {10.1016/j.asoc.2021.107492},
  urldate = {2024-11-15},
}

@inproceedings{ochoaStudyNKLandscapes2008,
  title = {A Study of {{NK}} Landscapes' Basins and Local Optima Networks},
  booktitle = {Proceedings of the 10th Annual Conference on {{Genetic}} and Evolutionary Computation},
  author = {Ochoa, Gabriela and Tomassini, Marco and V{\'e}rel, Seb{\'a}stien and Darabos, Christian},
  year = {2008},
  month = jul,
  series = {{{GECCO}} '08},
  pages = {555--562},
  publisher = {Association for Computing Machinery},
  address = {New York, NY, USA},
  doi = {10.1145/1389095.1389204},
  urldate = {2024-11-28},
  isbn = {978-1-60558-130-9},
}

@incollection{pitzerComprehensiveSurveyFitness2012,
  title = {A {{Comprehensive Survey}} on {{Fitness Landscape Analysis}}},
  booktitle = {Recent {{Advances}} in {{Intelligent Engineering Systems}}},
  author = {Pitzer, Erik and Affenzeller, Michael},
  editor = {Fodor, J{\'a}nos and Klempous, Ryszard and Su{\'a}rez Araujo, Carmen Paz},
  year = {2012},
  series = {Studies in {{Computational Intelligence}}},
  pages = {161--191},
  publisher = {Springer},
  address = {Berlin, Heidelberg},
  doi = {10.1007/978-3-642-23229-9_8},
  urldate = {2024-01-28},
  isbn = {978-3-642-23229-9},
  langid = {english},
}

@book{preussMultimodalOptimizationMeans2015,
  title = {Multimodal {{Optimization}} by {{Means}} of {{Evolutionary Algorithms}}},
  author = {Preuss, Mike},
  year = {2015},
  series = {Natural {{Computing Series}}},
  publisher = {Springer International Publishing},
  address = {Cham},
  doi = {10.1007/978-3-319-07407-8},
  urldate = {2024-07-02},
  copyright = {http://www.springer.com/tdm},
  isbn = {978-3-319-07406-1 978-3-319-07407-8},
}

@incollection{reevesFitnessLandscapes2014,
  title = {Fitness {{Landscapes}}},
  booktitle = {Search {{Methodologies}}: {{Introductory Tutorials}} in {{Optimization}} and {{Decision Support Techniques}}},
  author = {Reeves, Colin R.},
  editor = {Burke, Edmund K. and Kendall, Graham},
  year = {2014},
  pages = {681--705},
  publisher = {Springer US},
  address = {Boston, MA},
  doi = {10.1007/978-1-4614-6940-7_22},
  urldate = {2024-01-28},
  isbn = {978-1-4614-6940-7},
  langid = {english},
}

@article{reidysCombinatorialLandscapes2002,
  title = {Combinatorial {{Landscapes}}},
  author = {Reidys, Christian M. and Stadler, Peter F.},
  year = {2002},
  journal = {SIAM Review},
  volume = {44},
  number = {1},
  eprint = {4148412},
  eprinttype = {jstor},
  pages = {3--54},
  publisher = {{Society for Industrial and Applied Mathematics}},
  issn = {0036-1445},
  urldate = {2024-04-08},
}

@inproceedings{sanchez-diazRegularizedFeatureSelection2024,
  title = {Regularized {{Feature Selection Landscapes}}: {{An Empirical Study}} of~{{Multimodality}}},
  shorttitle = {Regularized {{Feature Selection Landscapes}}},
  booktitle = {Parallel {{Problem Solving}} from {{Nature}} -- {{PPSN XVIII}}},
  author = {{S{\'a}nchez-D{\'i}az}, Xavier F. C. and Masson, Corentin and Mengshoel, Ole Jakob},
  editor = {Affenzeller, Michael and Winkler, Stephan M. and Kononova, Anna V. and Trautmann, Heike and Tu{\v s}ar, Tea and Machado, Penousal and B{\"a}ck, Thomas},
  year = {2024},
  month = sep,
  pages = {409--426},
  publisher = {Springer Nature Switzerland},
  address = {Cham},
  doi = {10.1007/978-3-031-70055-2_25},
  copyright = {All rights reserved},
  isbn = {978-3-031-70055-2},
  langid = {english},
}

@misc{vanaretCertifiedGlobalMinima2020,
  title = {Certified {{Global Minima}} for a {{Benchmark}} of {{Difficult Optimization Problems}}},
  author = {Vanaret, Charlie and Gotteland, Jean-Baptiste and Durand, Nicolas and Alliot, Jean-Marc},
  year = {2020},
  month = mar,
  number = {arXiv:2003.09867},
  eprint = {2003.09867},
  primaryclass = {cs, math},
  publisher = {arXiv},
  doi = {10.48550/arXiv.2003.09867},
  urldate = {2023-05-25},
  archiveprefix = {arXiv},
}

@article{vanderlindenSurveyVisualizationTechniques2023,
  title = {A Survey of Visualization Techniques for Comparing Event Sequences},
  author = {{van der Linden}, Sanne and {de Fouw}, Evie and {van den Elzen}, Stef and Vilanova, Anna},
  year = {2023},
  month = oct,
  journal = {Computers \& Graphics},
  volume = {115},
  pages = {522--542},
  issn = {0097-8493},
  doi = {10.1016/j.cag.2023.05.016},
  urldate = {2024-11-23},
}

@phdthesis{wickhamPracticalToolsExploring2008,
  title = {Practical Tools for Exploring Data and Models},
  author = {Wickham, Hadley},
  year = {2008},
  school = {Iowa State University},
}

@book{wilkinsonGrammarGraphics2005,
  title = {The {{Grammar}} of {{Graphics}}},
  author = {Wilkinson, Leland},
  year = {2005},
  series = {Statistics and {{Computing}}},
  edition = {2},
  publisher = {Springer-Verlag},
  address = {New York},
  doi = {10.1007/0-387-28695-0},
  urldate = {2024-11-13},
  copyright = {http://www.springer.com/tdm},
  isbn = {978-0-387-24544-7},
  langid = {english},
}

@inproceedings{wrightRolesMutationInbreeding1932,
  title = {The Roles of Mutation, Inbreeding, Crossbreeding and Selection in Evolution, {{Proceedings}} of the {{Sixth International Congress}} of {{Genetics}}},
  booktitle = {Proc {{Sixth Int Congr Genet}}},
  author = {Wright, S.},
  year = {1932},
  volume = {1},
  pages = {356},
  urldate = {2024-11-28},
}

@inproceedings{yasudaAnalyzingViolationLandscapes2024,
  title = {Analyzing {{Violation Landscapes Using Different Definitions}} of {{Constraint Violation}}},
  booktitle = {Proceedings of the {{Genetic}} and {{Evolutionary Computation Conference Companion}}},
  author = {Yasuda, Yusuke and Tamura, Kenichi and Yasuda, Keiichiro},
  year = {2024},
  month = aug,
  series = {{{GECCO}} '24 {{Companion}}},
  pages = {1815--1823},
  publisher = {Association for Computing Machinery},
  address = {New York, NY, USA},
  doi = {10.1145/3638530.3664118},
  urldate = {2024-11-24},
  isbn = {979-8-4007-0495-6},
}

@misc{credit_approval_27,
  author       = {Quinlan, J. R.},
  title        = {{Credit Approval}},
  year         = {1987},
  howpublished = {UCI Machine Learning Repository},
  note         = {{DOI}: https://doi.org/10.24432/C5FS30}
}

@misc{ecoli_39,
  author       = {Nakai, Kenta},
  title        = {{Ecoli}},
  year         = {1996},
  howpublished = {UCI Machine Learning Repository},
  note         = {{DOI}: https://doi.org/10.24432/C5388M}
}

@misc{glass_identification_42,
  author       = {German, B.},
  title        = {{Glass Identification}},
  year         = {1987},
  howpublished = {UCI Machine Learning Repository},
  note         = {{DOI}: https://doi.org/10.24432/C5WW2P}
}

@article{gleicherConsiderationsVisualizingComparison2018,
  title = {Considerations for {{Visualizing Comparison}}},
  author = {Gleicher, Michael},
  year = {2018},
  month = jan,
  journal = {IEEE Transactions on Visualization and Computer Graphics},
  volume = {24},
  number = {1},
  pages = {413--423},
  issn = {1941-0506},
  doi = {10.1109/TVCG.2017.2744199},
  urldate = {2025-04-07},
}

@inproceedings{javedExploringDesignSpace2012,
  title = {Exploring the Design Space of Composite Visualization},
  booktitle = {2012 {{IEEE Pacific Visualization Symposium}}},
  author = {Javed, Waqas and Elmqvist, Niklas},
  year = {2012},
  month = feb,
  pages = {1--8},
  issn = {2165-8773},
  doi = {10.1109/PacificVis.2012.6183556},
  urldate = {2025-04-07},
}

@article{nobreStateArtVisualizing2019,
  title = {The {{State}} of the {{Art}} in {{Visualizing Multivariate Networks}}},
  author = {Nobre, C. and Meyer, M. and Streit, M. and Lex, A.},
  year = {2019},
  journal = {Computer Graphics Forum},
  volume = {38},
  number = {3},
  pages = {807--832},
  issn = {1467-8659},
  doi = {10.1111/cgf.13728},
  urldate = {2025-04-07},
  langid = {english},
}

@incollection{yolcu19Learning,
  title     = {Learning Local Search Heuristics for Boolean Satisfiability},
  author    = {Yolcu, E. and Poczos, B.},
  booktitle = {Proc. NeurIPS},
  pages     = {7992--8003},
  year      = {2019}
}

@inproceedings{schumann2020discrete,
  title     = {Discrete Optimization for Unsupervised Sentence Summarization with Word-Level Extraction},
  author    = {Schumann, R. and Mou, L. and Lu, Y. and Vechtomova, O. and Markert, K.},
  booktitle = {Proc. ACL},
  pages     = {5032--5042},
  year      = {2020}
}

@inproceedings{mengshoel2021stochastic,
  title     = {Stochastic Local Search Heuristics for Efficient Feature Selection: An Experimental Study},
  author    = {Mengshoel, O. J. and Flogard, E. and Riege, J.  and Yu, T.},
  booktitle = {Proc. NIKT},
  pages     = {58--71},
  year      = {2021}
}

@inproceedings{thomsonEntropySearchTrajectories2024,
  title = {Entropy, {{Search Trajectories}}, and~{{Explainability}} for~{{Frequency Fitness Assignment}}},
  booktitle = {Parallel {{Problem Solving}} from {{Nature}} -- {{PPSN XVIII}}},
  author = {Thomson, Sarah L. and Ochoa, Gabriela and {van den Berg}, Daan and Liang, Tianyu and Weise, Thomas},
  editor = {Affenzeller, Michael and Winkler, Stephan M. and Kononova, Anna V. and Trautmann, Heike and Tu{\v s}ar, Tea and Machado, Penousal and B{\"a}ck, Thomas},
  year = {2024},
  pages = {377--392},
  publisher = {Springer Nature Switzerland},
  address = {Cham},
  doi = {10.1007/978-3-031-70055-2_23},
  isbn = {978-3-031-70055-2},
}

@inproceedings{lara-cardenasImprovingHyperheuristicPerformance2019,
  title = {Improving {{Hyper-heuristic Performance}} for {{Job Shop Scheduling Problems Using Neural Networks}}},
  booktitle = {Advances in {{Soft Computing}}},
  author = {{Lara-C{\'a}rdenas}, Erick and {S{\'a}nchez-D{\'i}az}, Xavier and Amaya, Ivan and {Ortiz-Bayliss}, Jos{\'e} Carlos},
  editor = {{Mart{\'i}nez-Villase{\~n}or}, Lourdes and Batyrshin, Ildar and {Mar{\'i}n-Hern{\'a}ndez}, Antonio},
  year = {2019},
  pages = {150--161},
  publisher = {Springer International Publishing},
  address = {Cham},
  doi = {10.1007/978-3-030-33749-0_13},
  isbn = {978-3-030-33749-0},
}

@inproceedings{sanchez-diazPreliminaryStudyFeatureindependent2020,
  title = {A {{Preliminary Study}} on {{Feature-independent Hyper-heuristics}} for the 0/1 {{Knapsack Problem}}},
  booktitle = {2020 {{IEEE Congress}} on {{Evolutionary Computation}} ({{CEC}})},
  author = {{S{\'a}nchez-D{\'i}az}, Xavier F. C. and {Ortiz-Bayliss}, Jos{\'e} Carlos and Amaya, Ivan and {Cruz-Duarte}, Jorge M. and {Conant-Pablos}, Santiago Enrique and {Terashima-Mar{\'i}n}, Hugo},
  year = {2020},
  month = jul,
  pages = {1--8},
  doi = {10.1109/CEC48606.2020.9185671},
  urldate = {2023-11-30},
}

@article{sanchez-diazFeatureIndependentHyperHeuristicApproach2021,
  title = {A {{Feature-Independent Hyper-Heuristic Approach}} for {{Solving}} the {{Knapsack Problem}}},
  author = {{S{\'a}nchez-D{\'i}az}, Xavier and {Ortiz-Bayliss}, Jos{\'e} Carlos and Amaya, Ivan and {Cruz-Duarte}, Jorge M. and {Conant-Pablos}, Santiago Enrique and {Terashima-Mar{\'i}n}, Hugo},
  year = {2021},
  month = jan,
  journal = {Applied Sciences},
  volume = {11},
  number = {21},
  pages = {10209},
  publisher = {Multidisciplinary Digital Publishing Institute},
  issn = {2076-3417},
  doi = {10.3390/app112110209},
  urldate = {2022-11-18},
  copyright = {http://creativecommons.org/licenses/by/3.0/},
  langid = {english},
}

@article{zhangHybridAlgorithmVehicle2017,
  title = {A Hybrid Algorithm for a Vehicle Routing Problem with Realistic Constraints},
  author = {Zhang, Defu and Cai, Sifan and Ye, Furong and Si, Yain-Whar and Nguyen, Trung Thanh},
  year = {2017},
  month = jul,
  journal = {Information Sciences},
  volume = {394--395},
  pages = {167--182},
  issn = {0020-0255},
  doi = {10.1016/j.ins.2017.02.028},
  urldate = {2022-11-09},
  langid = {english},
}

@inproceedings{sanchez-diaz_estimating_2024,
    address = {New York, NY, USA},
    series = {{GECCO} '24 {Companion}},
    title = {Estimating the {Number} of {Local} {Optima} in {Multimodal} {Pseudo}-{Boolean} {Functions}: {Validation} via {Landscapes} of {Triangles}},
    copyright = {All rights reserved},
    isbn = {9798400704956},
    shorttitle = {Estimating the {Number} of {Local} {Optima} in {Multimodal} {Pseudo}-{Boolean} {Functions}},
    url = {https://dl.acm.org/doi/10.1145/3638530.3654156},
    doi = {10.1145/3638530.3654156},
    urldate = {2024-08-07},
    booktitle = {Proceedings of the {Genetic} and {Evolutionary} {Computation} {Conference} {Companion}},
    publisher = {Association for Computing Machinery},
    author = {Sánchez-Díaz, Xavier F. C. and Mengshoel, Ole Jakob},
    month = aug,
    year = {2024},
    pages = {211--214},
}

@inproceedings{antipov_local_2024,
    title = {Local {Optima} in {Diversity} {Optimization}: {Non}-trivial {Offspring} {Population} is {Essential}},
    isbn = {978-3-031-70071-2},
    shorttitle = {Local {Optima} in {Diversity} {Optimization}},
    url = {https://link.springer.com/chapter/10.1007/978-3-031-70071-2_12},
    doi = {10.1007/978-3-031-70071-2_12},
    language = {en},
    urldate = {2024-09-17},
    booktitle = {Parallel {Problem} {Solving} from {Nature} – {PPSN} {XVIII}},
    publisher = {Springer, Cham},
    author = {Antipov, Denis and Neumann, Aneta and Neumann, Frank},
    year = {2024},
    note = {ISSN: 1611-3349},
    pages = {181--196},
}

@article{dang_populations_2017,
    title = {Populations {Can} {Be} {Essential} in {Tracking} {Dynamic} {Optima}},
    volume = {78},
    issn = {1432-0541},
    url = {https://doi.org/10.1007/s00453-016-0187-y},
    doi = {10.1007/s00453-016-0187-y},
    language = {en},
    number = {2},
    urldate = {2022-07-13},
    journal = {Algorithmica},
    author = {Dang, Duc-Cuong and Jansen, Thomas and Lehre, Per Kristian},
    month = jun,
    year = {2017},
    pages = {660--680},
}

@inproceedings{mengshoel_understanding_2022,
    address = {New York, NY, USA},
    series = {{GECCO} '22},
    title = {Understanding the cost of fitness evaluation for subset selection: {Markov} chain analysis of stochastic local search},
    isbn = {978-1-4503-9237-2},
    shorttitle = {Understanding the cost of fitness evaluation for subset selection},
    url = {https://doi.org/10.1145/3512290.3528689},
    doi = {10.1145/3512290.3528689},
    urldate = {2022-07-14},
    booktitle = {Proceedings of the {Genetic} and {Evolutionary} {Computation} {Conference}},
    publisher = {Association for Computing Machinery},
    author = {Mengshoel, Ole Jakob and Flogard, Eirik Lund and Yu, Tong and Riege, Jon},
    month = jul,
    year = {2022},
    pages = {251--259},
}

@inproceedings{schneider_hpo_2022,
    address = {Cham},
    series = {Lecture {Notes} in {Computer} {Science}},
    title = {{HPO} $\times$ {ELA}: {Investigating} {Hyperparameter} {Optimization} {Landscapes} by {Means} of {Exploratory} {Landscape} {Analysis}},
    isbn = {978-3-031-14714-2},
    shorttitle = {{HPO} \$\${\textbackslash}times \$\${ELA}},
    doi = {10.1007/978-3-031-14714-2_40},
    language = {en},
    booktitle = {Parallel {Problem} {Solving} from {Nature} – {PPSN} {XVII}},
    publisher = {Springer International Publishing},
    author = {Schneider, Lennart and Schäpermeier, Lennart and Prager, Raphael Patrick and Bischl, Bernd and Trautmann, Heike and Kerschke, Pascal},
    editor = {Rudolph, Günter and Kononova, Anna V. and Aguirre, Hernán and Kerschke, Pascal and Ochoa, Gabriela and Tušar, Tea},
    year = {2022},
    pages = {575--589},
}

@inproceedings{lu_nsga-net_2019,
    address = {New York, NY, USA},
    series = {{GECCO} '19},
    title = {{NSGA}-{Net}: neural architecture search using multi-objective genetic algorithm},
    isbn = {978-1-4503-6111-8},
    shorttitle = {{NSGA}-{Net}},
    url = {https://doi.org/10.1145/3321707.3321729},
    doi = {10.1145/3321707.3321729},
    urldate = {2023-09-01},
    booktitle = {Proceedings of the {Genetic} and {Evolutionary} {Computation} {Conference}},
    publisher = {Association for Computing Machinery},
    author = {Lu, Zhichao and Whalen, Ian and Boddeti, Vishnu and Dhebar, Yashesh and Deb, Kalyanmoy and Goodman, Erik and Banzhaf, Wolfgang},
    month = jul,
    year = {2019},
    pages = {419--427},
}

@article{stanley_designing_2019,
    title = {Designing neural networks through neuroevolution},
    volume = {1},
    copyright = {2019 Springer Nature Limited},
    issn = {2522-5839},
    url = {https://www.nature.com/articles/s42256-018-0006-z},
    doi = {10.1038/s42256-018-0006-z},
    language = {en},
    number = {1},
    urldate = {2022-09-27},
    journal = {Nature Machine Intelligence},
    author = {Stanley, Kenneth O. and Clune, Jeff and Lehman, Joel and Miikkulainen, Risto},
    month = jan,
    year = {2019},
    note = {Number: 1
Publisher: Nature Publishing Group},
    pages = {24--35},
}

@misc{heart_disease_45,
  author       = {Janosi, Andras and Steinbrunn, William and Pfisterer, Matthias and Detrano, Robert},
  title        = {{Heart Disease}},
  year         = {1989},
  howpublished = {UCI Machine Learning Repository},
  note         = {{DOI}: https://doi.org/10.24432/C52P4X}
}

@article{mostert_feature_2021,
    title = {A {Feature} {Selection} {Algorithm} {Performance} {Metric} for {Comparative} {Analysis}},
    volume = {14},
    copyright = {http://creativecommons.org/licenses/by/3.0/},
    issn = {1999-4893},
    url = {https://www.mdpi.com/1999-4893/14/3/100},
    doi = {10.3390/a14030100},
    language = {en},
    number = {3},
    urldate = {2024-01-28},
    journal = {Algorithms},
    author = {Mostert, Werner and Malan, Katherine M. and Engelbrecht, Andries P.},
    month = mar,
    year = {2021},
    note = {Number: 3
Publisher: Multidisciplinary Digital Publishing Institute},
    pages = {100},
}

@misc{thomson_stalling_2024,
    title = {Stalling in {Space}: {Attractor} {Analysis} for any {Algorithm}},
    shorttitle = {Stalling in {Space}},
    url = {http://arxiv.org/abs/2412.15848},
    doi = {10.48550/arXiv.2412.15848},
    urldate = {2025-04-03},
    publisher = {arXiv},
    author = {Thomson, Sarah L. and Renau, Quentin and Vermetten, Diederick and Hart, Emma and Stein, Niki van and Kononova, Anna V.},
    month = dec,
    year = {2024},
    note = {arXiv:2412.15848 [cs]},
}

\end{document}